\documentclass[12pt,a4paper]{article}
\usepackage[T2A]{fontenc}
\usepackage{amsfonts}
\usepackage{amsmath}
\usepackage{amssymb}
\usepackage[english]{babel}
\usepackage{graphicx}
\usepackage{wrapfig}
\usepackage{caption}
\begin{document}
\title{An example of the stochastic dynamics of a causal set}

\author{Alexey L. Krugly\thanks{Scientific Research Institute for System Analysis of the Russian Academy of Science, 117218, Nahimovskiy pr., 36, k. 1, Moscow, Russia; akrugly@mail.ru.}\and Ivan V. Stepanian\thanks{Mechanical Engineering Research Institute (Department of Biomechanics) of the Russian Academy of Sciences,
101990, Malyj Hariton'evskij per., 4, Moscow, Russia; skwwwks@gmail.com.}}
\date{} \maketitle
\begin{abstract}
An example of a discrete pregeometry on a microscopic scale is introduced. The model is a directed dyadic acyclic graph. This is the particular case of a causal set. The particles in this model must be self-organized repetitive structures. The dynamics of this model is a stochastic sequential growth dynamics. New vertexes are added one by one. The probability of this addition depends on the structure of existed graph. The particular case of the dynamics is considered. The numerical simulation provides some symptoms of self-organization.


\end{abstract}

We consider a model of a discrete pregeometry. By assumption there are not spacetime and matter on a microscopic scale. There are only a discrete pregeometry. Usually physics describes matter in spacetime. Otherwise the far goal of this model is to describe particles as symmetrical structures of pregeometry without any reference to continuous spacetime. We suppose that continuous spacetime emerges only as a property of particles if the number of particles tends to infinity.

\begin{wrapfigure}[7]{r}{6cm}
	\centering	
				\includegraphics[trim=0cm 0cm 0cm 0cm]{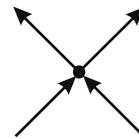}
		\captionsetup{labelformat=simple,font=footnotesize,labelsep=period}
	\caption{An x-structure.}
	\label{fig:fig1}
\end{wrapfigure}

The model of the universe is an infinite directed dyadic acyclic graph. The directed graph means that all edges are directed. The dyadic graph means that each vertex has two incident incoming edges and two incident outgoing edges. The vertex with incident edges forms an x-structure (Fig.\ \ref{fig:fig1}). The acyclic graph means that there is not a directed loop. Hereinafter only such graph is considered and it is called a graph for short. This model was introduced by D. Finkelstein \cite{Fink88} in 1988. This model is a particular case of a causal set. A causal set approach to quantum gravity was introduced by G. 't Hooft \cite{'t Hooft} and J. Myrheim \cite{Myrheim} in 1978. A causal set is a partially ordered locally finite set (see the review \cite{Sorkin2005}).

The model of the universe is an infinite graph. But any observer can only actually know a finite number of facts.  Then we consider only finite graphs. In a graph theory, by definition, an edge is a relation of two vertexes.
Consequently some vertexes of finite graph have less than four incident edges. These vertexes have free valences instead the absent edges. These free valences are called external edges as external lines in Feynman diagrams. They are figured as edges that are incident to only one vertex. There are incoming and outgoing external edges. We can prove that the number of incoming external edges is equal to the number of outgoing external edges for any such graph \cite[Lemma~5]{1008.5169}. 

Each such graph is a model of a part of some process. The task is to predict the future stages of this process or to reconstruct the past stages. We can reconstruct the graph step by step. The minimal part is a vertex. We start from some given graph and add new vertexes one by one. This procedure is proposed by A. L. Krugly \cite{Krugly1998,Krugly2002} in 1998. Similar procedure and the term `a classical sequential growth dynamics' are proposed by D. P. Rideout and R. D. Sorkin \cite{RideoutSorkin} for other model of a causal set in 1999.

\begin{figure}[t]
	\centering	
		\includegraphics[width=13.5cm,trim=0cm 0cm 0cm 0cm]{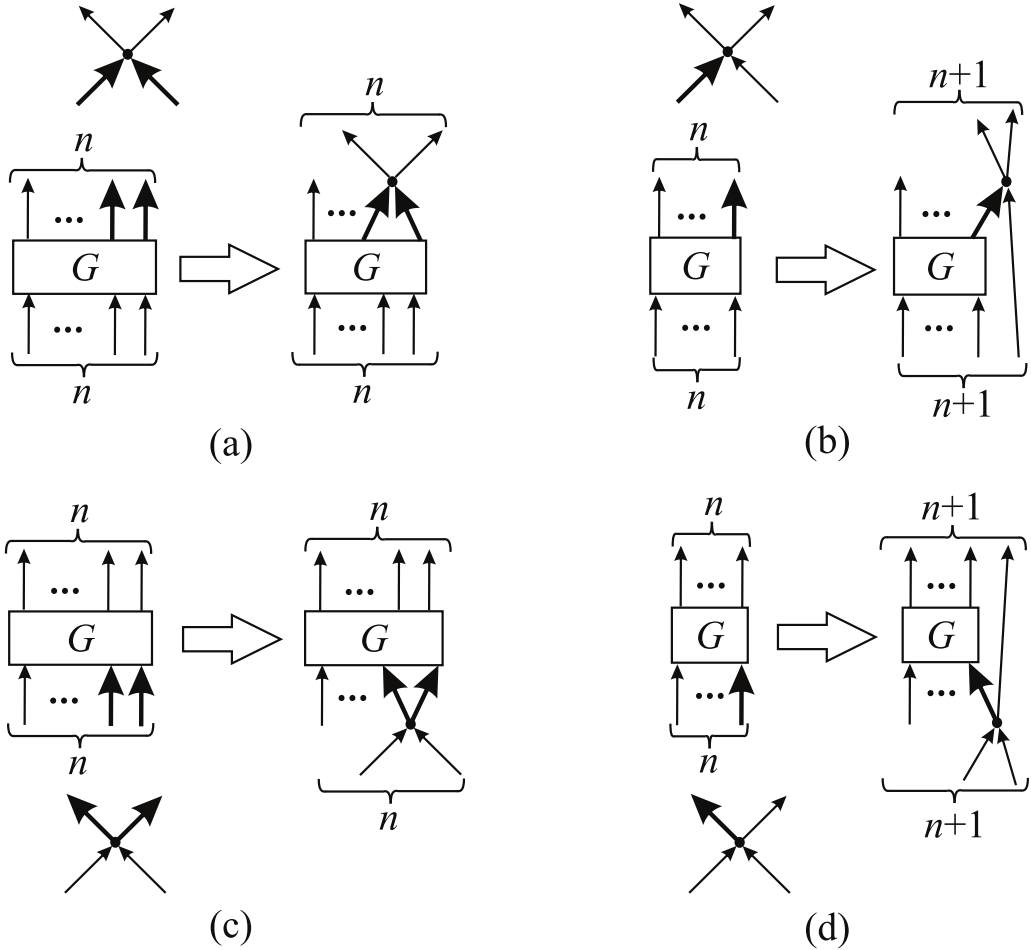}
	\caption{The types of elementary extensions: (a) the first type, (b) the second type, (c) the third type, and (d) the fourth type.}
	\label{fig:fig2}
\end{figure}

We can add a new vertex to external edges. This procedure is called an elementary extension. There are four types of elementary extensions \cite{1004.5077}. There are two types of elementary extensions to outgoing external edges (Fig.\ \ref{fig:fig2} (a) and\ \ref{fig:fig2} (b)). This is a reconstruction of the future of the process.  In this and following figures the graph $G$ is represented by a rectangle because it can have an arbitrary structure. The edges that take part in the elementary extension are figured by bold arrows. First type is an elementary extension to two outgoing external edges (Fig.\ \ref{fig:fig2} (a)). Second type is an elementary extension to one outgoing external edge (Fig.\ \ref{fig:fig2} (b)).
Similarly, there are two types of elementary extensions to incoming external edges (Fig.\ \ref{fig:fig2} (c) and\ \ref{fig:fig2} (d)). These elementary extensions reconstruct the past evolution of the process. Third type is an elementary extension to two incoming external edges (Fig.\ \ref{fig:fig2} (c)). Fourth type is an elementary extension to one incoming external edge (Fig.\ \ref{fig:fig2} (d)). We can prove that we can get every connected graph by a sequence of elementary extensions of these four types \cite[Teorem~2]{1008.5169}.

By assumption, the dynamics of this model is a stochastic dynamics. We can only calculate probabilities of different variants of elementary extensions. The algorithm of calculation of probabilities includes 3 steps \cite{1106.6269}. The first step is the choice of the elementary extension to the future or to the past. We assume the probability of this choice is $1/2$. The second step is the choice of the first external edge that takes part in the elementary extension. We assume the equiprobable choice. The probability of this choice is $1/n$. The third step is the choice of second external edge that takes part in the elementary extension. In general case, these probabilities depend on the structure of the existed graph. By assumption, the calculation of these probabilities is based on binary alternatives.

\begin{figure}[t]
	\centering	
		\includegraphics[width=10cm,trim=0cm 0cm 0cm 0cm]{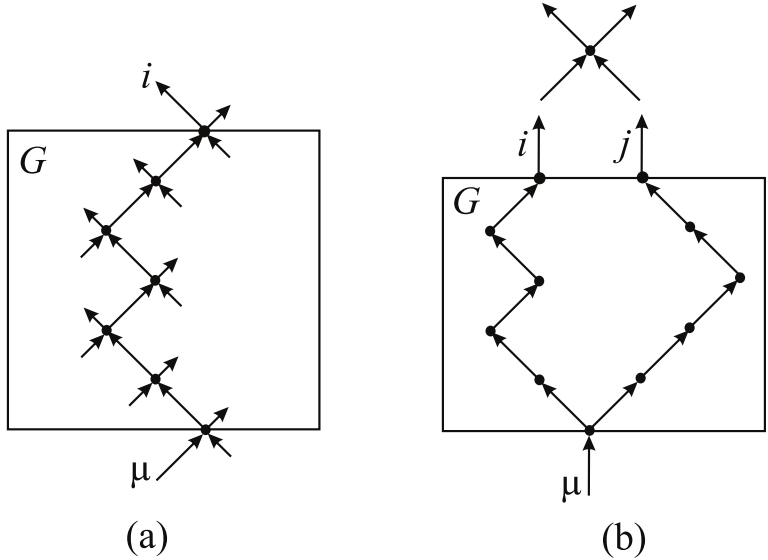}
	\caption{(a) A choice of a directed path is a sequence of binary alternatives. (b) A new loop is generated by a new vertex.}
	\label{fig:fig3}
\end{figure}

Number outgoing external edges by Latin indices. Number incoming external edges by Greek indices. Choose a directed path from any incoming external edge number $\mu$ (Fig.\ \ref{fig:fig3} (a)). This path ends in some outgoing external edge number $i$. We must choose 1 edge in each vertex. We assume the equiprobable choice. Consequently if a directed path includes $k$ vertexes, the choice of this path has the probability $2^{-k}$.
Introduce an amplitude $a_{i\mu}$ of the causal connection of the incoming external edge number $\mu$ and the outgoing external edge number $i$. By definition, this is a sum of probabilities of all directed paths between these edges.
\begin{equation}
\label{eq:1} a_{i\mu}= a_{\mu i}=\sum_{m=1}^M 2^{-k(m)}\textrm{,}
\end{equation}
where $M$ is the number of all directed paths between these edges, $k(m)$ is the number of vertexes in the path number $m$. This is not a complex quantum amplitude. This is real nonnegative number. This definition has clear physical meaning. The causal connection of two edges is stronger if there are more directed paths between these edges and these paths are shorter. The addition of a new vertex to two external edges forms a set of loops. Each loop is formed by two directed paths (Fig.\ \ref{fig:fig3} (b)). We assume that the probability to generate this loop is a product of probabilities of these paths. By assumption, the probability of an elementary extension is directly proportional to the sum of probabilities of new loops that are generated by this elementary extension. By definition, put
\begin{equation}
\label{eq:2} p_{ij}=\sum_{\mu=1}^n a_{i\mu} a_{\mu j}\textrm{.}
\end{equation}
$p_{ij}$ is the probability to add a new vertex to the outgoing external edges numbers $i$ and $j$. Similarly, $p_{\alpha\beta}$ is the probability to add a new vertex to the incoming external edges numbers $\alpha$ and $\beta$.
\begin{equation}
\label{eq:3} p_{\alpha \beta}=\sum_{i=1}^n a_{\alpha i} a_{i\beta}\textrm{,}
\end{equation}
The third step of the algorithm is the choice of second external edge with these probabilities. We get the right normalization if the probability of the addition to 1 external edge is this probability where $i=j$ ($\alpha=\beta$). In this case, the external edges numbers $i$ and $j$ ($\alpha$ and $\beta$) coincide. We can get evolution equations for amplitudes during sequential growth of a graph \cite{1106.6269}. We have the quadratic equations (\ref{eq:2}) - (\ref{eq:3}) for probabilities. This is like quantum theory. But in this model, all numbers are real and nonnegative.

\begin{figure}[t]
	\centering	
		\includegraphics[width=14.7cm,trim=0cm 0cm 0cm 0cm]{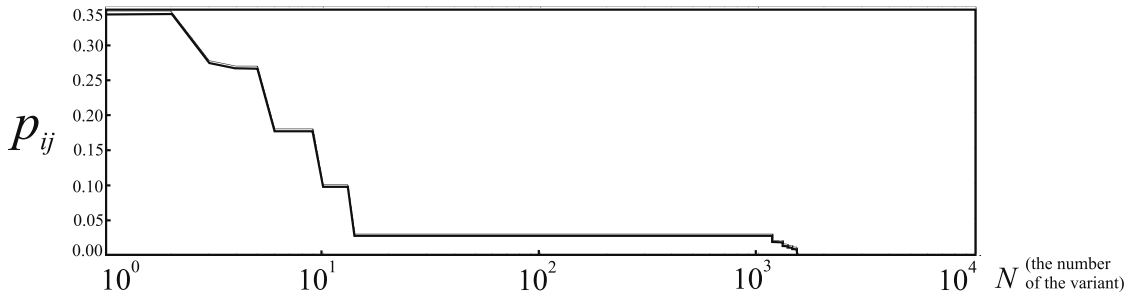}
	\caption{The probabilities of different variants to add a new vertex to the future in the step number 500.}
	\label{fig:fig4}
\end{figure}

\begin{figure}[t]
	\centering
		\includegraphics[width=6.5cm,trim=7cm 17cm 7cm 4cm]{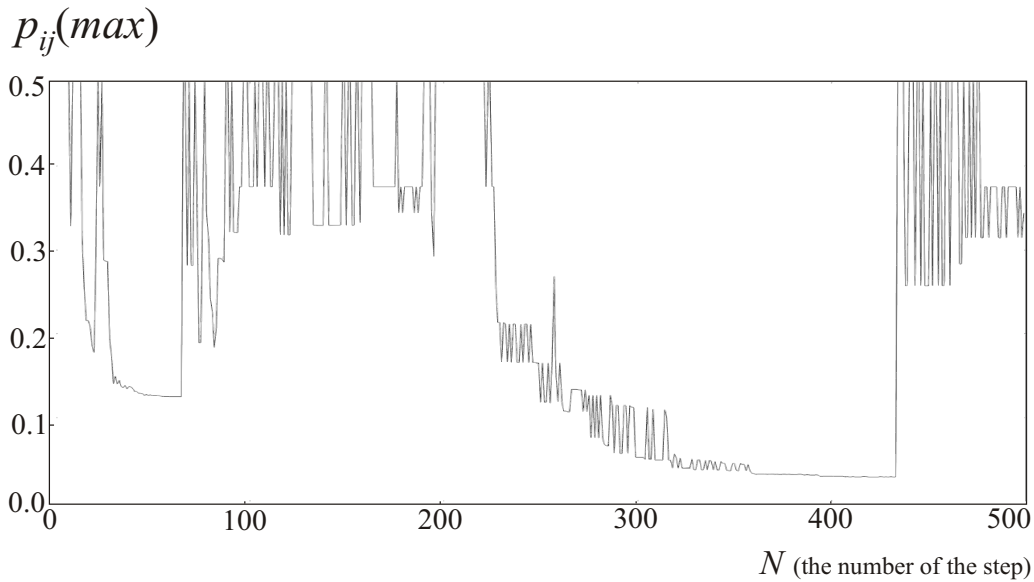}
	\caption{The maximal probability in each step of sequential grows.}
	\label{fig:fig5}
\end{figure}

This model is convenient for numerical simulation. Now we begin this investigation. Consider one example. We start from 1 vertex and calculate 500 steps. Consider probabilities of different variants to add a new vertex to the future in the step number 500 (Fig.\ \ref{fig:fig4}).
These probabilities are ranged from maximal to minimum value. A log scale of the numbers of variants is used. There are 1521 variants. We see that there are about 10 variants with high probability. These are preferable variants of the growth. Consider the maximal probability in each step (Fig.\ \ref{fig:fig5}).
This probability aperiodically oscillates during sequential growth. The maximal possible value is $1/2$. We see that there are a variant with high probability in many steps. We hope that the existence of the small quantity of preferable variants of the growth is a symptom of self-organization.
We hope that such structures describe particles. This self-organization is a task for our further investigation. We are going to calculate main frequencies of the dynamics. In the truth model, ratios of main frequencies must be the same as mass ratios of real particles.

\end{document}